\documentclass[aps,prl,reprint,groupedaddress,footinbib]{revtex4-2}

\usepackage{graphicx}
\usepackage{verbatim}
\usepackage{amsfonts} 
\usepackage{amsmath}
\usepackage{mathbbol}
\usepackage{amssymb}

\begin{document}


\title{Multiple Scattering Expansion for Dielectric Media: Casimir Effect}


\author{Thorsten Emig}
\affiliation{Laboratoire de Physique
Th\'eorique et Mod\`eles Statistiques, CNRS UMR 8626, B\^at.~530,
Universit\'e Paris-Sud/Saclay, 91405 Orsay cedex, France}

\author{Giuseppe Bimonte}
\affiliation{Dipartimento di Fisica E. Pancini, Università di Napoli Federico II,
Complesso Universitario di Monte S. Angelo,
Via Cinthia, I-80126 Napoli, Italy}
\affiliation{INFN Sezione di Napoli, I-80126 Napoli, Italy}


\date{\today}

\begin{abstract} 
Recent measurements of Casimir forces have provided evidence of an
intricate modification of quantum fluctuations of the electromagnetic
field in complex geometries.
Here we introduce a multiple scattering description for
Casimir interactions between bodies of arbitrary shape and material
composition, admitting an expansion as a sequence of inter- and intra-body
 wave scatterings.  Interactions in complex
geometries can be computed within current experimental
resolution from typically a few wave scatterings, notably without any
a-priori knowledge of the scattering amplitudes of the bodies.
Some first applications demonstrate the power of the approach.
\end{abstract}


\maketitle




{\it Introduction} --- A striking feature of quantum electrodynamics is that quantum fluctuations of the electromagnetic (EM) field can interact with matter to produce measurable long range interactions between atomic particles and consequently macroscopic bodies. A spectacular manifestation of this interaction is the attractive force between two perfectly conducting plates predicted by Casimir \cite{casimir}. 
The connection between an atomistic description and non-ideal macroscopic dielectric materials was established by Lifshitz \cite{lifshitz}.  In the past decades, this approach has been the core theory for interpreting most of the precision measurements of Casimir interactions \cite{lamoreaux,mohideen,chan,bressi,Decca:2003yb,Munday:2009xw,Sushkov:2011ik,Tang:2017kz,Bimonte:2016cr}.  
Importantly, measurements almost always involve bodies with non-planar surfaces, either to avoid difficulties in parallel alignment or to intentionally detect geometry effects. Consequently, Lifshitz's planar surface model is  combined with the proximity force approximation (PFA) to obtain a force estimate by decomposing the surfaces into pairs of small and parallel patches \cite{Derjaguin:1934hb}. A breakdown of this approach is expected with increasing separation as it ignores complex diffraction effects and the non-additivity of Casimir forces. Indeed, recent experiments have demonstrated large deviations from PFA \cite{Banishev:2013zp,Intravaia:2013yf,Wang2021}, making theoretical formulations for a precise force computation highly desirable. Such computational method is relevant to a plethora of fields like chemistry, structural biology, colloidal science and atomic physics.

To date, enormous efforts have been put forward by many groups to develop theoretical and numerical methods that can cope with more general surface shapes \cite{Rodriguez:2011df,bimonte2017}.  Specifically, the scattering method \cite{emig2007,kenneth2008,rahi2009}, originally devised for mirrors \cite{Genet03,lambrecht}, expresses the interaction between dielectric bodies in terms of their scattering amplitude, known as T-operator. While this approach has enabled most of recent theoretical progress, the T-operator is known only for highly symmetric bodies, such as sphere and cylinder, or for a few perfectly conducting shapes \cite{maghrebi}, practically exhausting this method. A more fundamental limitation is that interlocked geometries evade this method due to lack of convergence of the mode expansion \cite{Wang2021}.  The theoretical treatment of non-ideal materials with sharp surface features is beyond the scope of existing methods. A notable exception is a more recent, powerful fully numerical scheme based on a boundary element method (SCUFF-EM) \cite{reid2013,rodriguez,scuff}. It is believed that this approach can provide in principle the exact force for arbitrary shapes, with computational power the only but practically important limiting factor \cite{Wang2021}. To our knowledge, complementary, not fully numerical methods with comparably broad application range do not exist to date.

The main aim of our work is to overcome the difficulties residing in the existing scattering method: The lack of knowledge of T-operators for arbitrary shapes and the limited convergence of the corresponding partial wave expansions. The idea of our approach is to describe the fluctuating EM field in the presence of material bodies by treating the back and forth scatterings of waves between different objects on an equal footing as the scatterings within an isolated object, without making use of the concept of a T-operator. This gives rise to an expansion in the number of scattering points, where each scattering is treated exactly without resorting to any expansion in a partial wave basis. 
In fact, in a seminal work Balian and Duplantier demonstrated the very existence and convergence of a multiple scattering expansion (MSE) for Casimir forces which however is limited to perfect conductors \cite{balian1977,balian1978}.
Here we demonstrate the existence of a MSE for Casimir interactions for arbitrary dissipative magneto-dielectric materials.  

\begin{figure}
\centering
\includegraphics[width=1\linewidth]{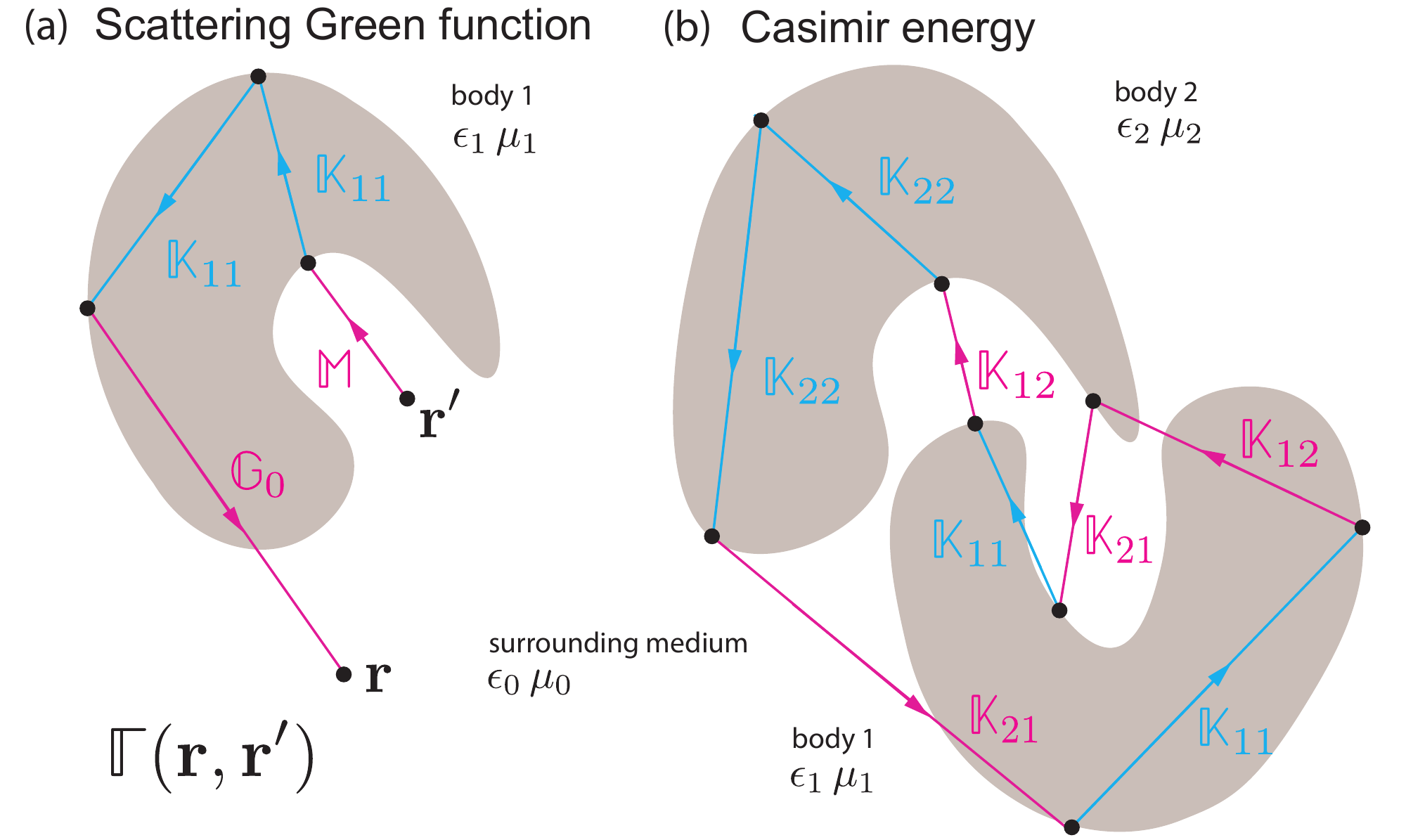}
\caption{\label{fig:1} Diagrammatic representation of contributions to the MSE, shown in panel {\bf (a)} for the scattering Green  function $\mathbb{\Gamma}({\bf r},{\bf r}')$ of a single body with source point ${\bf r}'$ and observation point ${\bf r}$, and in panel {\bf (b)} for the Casimir energy between two bodies. In the displayed examples, lines with arrows represent free propagation between surface points of the same body (blue lines) and to external points or between surface points of different bodies (magenta lines). See text for details.}
\end{figure}


{\it Multiple scattering expansion} --- We consider Casimir forces in a configuration of $N$ material bodies with dielectric and magnetic permittivities $\epsilon_\sigma$ and $\mu_\sigma$ ($\sigma=1,\ldots,N$) in thermal equilibrium at temperature $T$. The bodies are bounded by surfaces $S_\sigma$ which can be of arbitrary shape and separate their bulk from the surrounding homogeneous medium with dielectric and magnetic permittivities $\epsilon_0$ and $\mu_0$.
From the uniqueness of an EM field in a region specified by sources within the region and the tangential components of the field over the boundary of the region, one can construct the  EM field  $({\bf E}, {\bf H})$ separately in the region external to the bodies $({\bf E}^{(0)},{\bf H}^{(0)})$,  and in the $N$ interior regions of the bodies $({\bf E}^{(\sigma)},{\bf H}^{(\sigma)})$.  We introduce the surface currents ${\bf j}_{\sigma-}=-{\bf n}_\sigma \times {\bf H}_-$, ${\bf m}_{\sigma-}={\bf n}_\sigma \times {\bf E}_-$ as the tangential fields when $S_\sigma$ is approached from the inside of the bodies, and ${\bf j}_{\sigma+}={\bf n}_\sigma \times {\bf H}_+$, ${\bf m}_{\sigma+}=-{\bf n}_\sigma \times {\bf E}_+$ as the tangential fields when $S_\sigma$ is approached from the outside of the bodies. Here ${\bf n}_\sigma$ denotes the outward surface normal on surface $S_\sigma$. Further we assume that the incident field is generated by electric and magnetic currents $({\bf J}, {\bf M})$ outside the bodies.
Then the field in the interior of the bodies can be expressed  as the surface integral $({\bf E}^{(\sigma)},{\bf H}^{(\sigma)})({\bf r})= \int_{S_\sigma} ds_{{\bf u}} \, \mathbb{G}_\sigma({\bf r},{\bf u}) ({\bf j}_{\sigma-},{\bf m}_{\sigma-})({\bf u})$ where $\mathbb{G}_\sigma$ is the free Green tensor in a medium with permittivities $\epsilon_\sigma$, $\mu_\sigma$. Similarly, 
exterior to the bodies the field $({\bf E}^{(0)},{\bf H}^{(0)})({\bf r})= \int d{\bf r}' \, \mathbb{G}_0({\bf r},{\bf r}') ({\bf J},{\bf M})({\bf r}') + \sum_{\sigma=1}^N \int_{S_\sigma} ds_{{\bf u}} \, \mathbb{G}_0({\bf r},{\bf u}) ({\bf j}_{\sigma+},{\bf m}_{\sigma+})({\bf u})$. It can be shown \cite{muller} that the currents can be expressed in terms of the fields {\it on} the surface,  $({\bf m}_{\sigma-},-{\bf j}_{\sigma-}) ({\bf u}) = 2 {\bf n}_\sigma({\bf u}) \times ({\bf E}^{(\sigma)},{\bf H}^{(\sigma)})({\bf u})$ and $({\bf m}_{\sigma+},-{\bf j}_{\sigma+})({\bf u}) = - 2 {\bf n}_\sigma({\bf u}) \times ({\bf E}^{(0)},{\bf H}^{(0)})({\bf u})$ for ${\bf u}$ located on surface $S_\sigma$.
A consistent set of $2N$ integral equations in the unknown surface currents $({\bf j}_{\sigma},{\bf m}_{\sigma})\equiv ({\bf j}_{\sigma+},{\bf m}_{\sigma+}) =- ({\bf j}_{\sigma-},{\bf m}_{\sigma-})$ with a unique solution can be obtained  by imposing the usual continuity conditions on the tangential field components at the interfaces between different media,
and by taking linear combinations of the set of $2N$ equations involving $({\bf E}^{(\sigma)},{\bf H}^{(\sigma)})$ {\it and} the corresponding set involving $({\bf E}^{(0)},{\bf H}^{(0)})$ but not by considering only one of the two sets as this would ignore the coupling of the interior and exterior fields \cite{harrington}. In general, one can choose $4N$ suitable coefficients which form $2N$ diagonal $2\times 2$ matrices $\mathbb{C}^{i}_\sigma$, $\mathbb{C}^{e}_\sigma$ acting on the two field components of the interior and exterior integral equations. To interpret the integral equations as successive scatterings, we introduce the surface scattering operators (SSOs) $\mathbb{K}_{\sigma\sigma'}({\bf u},{\bf u}')$ which describe free propagation from ${\bf u}'$ on surface $S_{\sigma'}$ to ${\bf u}$ on surface $S_{\sigma}$ and scattering at point ${\bf u}$ 
\begin{eqnarray}
\!\!\!\!&&\mathbb{K}_{\sigma\sigma'}({\bf u},{\bf u}') = 2 \mathbb{P} (\mathbb{C}^{i}_\sigma+\mathbb{C}^{e}_\sigma)^{-1} {\bf n}_\sigma({\bf u})\nonumber \\
\label{eq:1}
\!\!\!\!&&\times  \left[  \delta_{\sigma\sigma'} \mathbb{C}^{i}_\sigma  \mathbb{G}_\sigma({\bf u},{\bf u}')- \mathbb{C}^{e}_\sigma \mathbb{G}_0({\bf u},{\bf u}')\right] \, , \quad \mathbb{P} = \big(\begin{smallmatrix} 0 & -1 \\ 1 & 0 \end{smallmatrix} \big)
\end{eqnarray}
acting on electric and magnetic tangential surface fields at ${\bf u}'$ ($\delta_{\sigma\sigma'}$ is the Kronecker delta).
 With these SSOs the surface currents are determined by the Fredholm integral equations
\begin{eqnarray}
\label{eq:2}
&&\sum_{\sigma'=1}^N\int_{S_{\sigma'}}\!\!ds_{{\bf u}'} \,\left[\mathbb{1} - \mathbb{K}_{\sigma\sigma'}({\bf u},{\bf u}')\right] \big(\begin{smallmatrix} {\bf j}_{\sigma'} \\ {\bf m}_{\sigma'} \end{smallmatrix} \big) ({\bf u}') \nonumber \\
&&= \!\int d{\bf r} \,\mathbb{M}_{\sigma}({\bf u},{\bf r})\big(\begin{smallmatrix} {\bf J} \\ {\bf M}\end{smallmatrix}\big) ({\bf r})  
\quad \text{with} \\
\mathbb{M}_\sigma({\bf u},{\bf r}) && =  -2 \mathbb{P} (\mathbb{C}^{i}_\sigma+\mathbb{C}^{e}_\sigma)^{-1} \mathbb{C}^{e}_\sigma\, {\bf n}_\sigma({\bf u}) \times  \mathbb{G}_0({\bf u},{\bf r}) \, . \nonumber
\end{eqnarray}

It can be shown  that the Casimir force on body $\sigma$ is determined solely by the SSO, expressed as a sum over Matsubara frequencies $\xi_n$ by  ${\bf F}_\sigma =  k_B T \sideset{}{'}\sum_{n=0}^\infty {\rm Tr} [ (\mathbb{1}-\mathbb{K})^{-1} \nabla_{{\bf r}_\sigma} {\mathbb K}]$ where $\nabla_{{\bf r}_\sigma}$ is the gradient with respect to the position of the body, the primed sum gives a weight of $1/2$ to the $n=0$ term, and the trace Tr involves a sum over vector indices of the electric and magnetic components and an integration over all surfaces (for details on the derivation see \cite{bimonte2021}).
To gain insight into the structure of the MSE for the Casimir energy, we consider two bodies. After subtracting the self-energies, arising from isolated scatterings on a single body, the energy is expressed in terms of four SSO as
\begin{equation}
\label{eq:4}
{\cal E} = k_B T \sideset{}{'}\sum_{n=0}^\infty {\rm Tr} \log \left[ \mathbb{1} - (\mathbb{1}-\mathbb{K}_{11})^{-1} \mathbb{K}_{12}(\mathbb{1}-\mathbb{K}_{22})^{-1}\mathbb{K}_{21}\right] \, .
\end{equation}
We note that the above formulae provide an exact representation of the EM field and the Casimir energy for arbitrary boundary shapes, independent of the allowed choices of the coefficients $\mathbb{C}^{i}_\sigma$, $\mathbb{C}^{e}_\sigma$ (see also next section). 
After expanding both the logarithm and the inverse operators in powers of the SSOs we obtain the MSE which involves at least one scattering on each body with closed paths going from body 1 to body 2 and back ($\mathbb{K}_{12}$ and $\mathbb{K}_{12}$), possibly multiple times, and with an arbitrary number (including zero) of scatterings on each body ($\mathbb{K}_{11}$ and $\mathbb{K}_{22}$), as illustrated in Fig.~\ref{fig:1}(b). 
We note that compared to the Born series, which is an expansion in terms of iterated integrals over the {\it volumes} occupied by the bodies  \cite{buhmann2}, our MSE in an expansion in terms of iterated integrals over the bodies {\it surfaces}. This property implies an enormous reduction of computing time, especially when high orders are considered.  While the Born series is an expansion in the dielectric contrast, our MSE is an expansion in the number of scatterings. 

{\it Equivalent formulations} --- With different  coefficient matrices $\mathbb{C}^{i/e}_\sigma$  the SSO form an equivalence class of operators in the sense that Eq.~(\ref{eq:2}) yields the same surface currents for a given external source, as long as neither the interior nor the exterior matrices vanish for any $\sigma$, and the sum $\mathbb{C}^{i}_\sigma+\mathbb{C}^{e}_\sigma$ is invertible. 
The surface currents and the Casimir energy at any {\it finite} order of the MSE, however, in general do depend on the chosen coefficients, and hence does the rate of convergence of the MSE. This remarkable property provides an effective method to optimize convergence for different permittivities and even frequencies by suitable adjustment of coefficients. 
Among the infinitely many choices there are a few which we consider important to discuss explicitly: (C1) In general, the SSO has a leading singularity that diverges as $1/|{\bf u}-{\bf u}'|^\gamma$ with $\gamma=3$ when the two surface positions ${\bf u}$, ${\bf u}'$ approach each other. 
There exists a choice of coefficients \cite{muller}, however, for which the singularity is reduced to a weaker divergence with exponent $\gamma=1$, presumably accelerating convergence. The coefficients are $\mathbb{C}^{i}_\sigma={\rm diag}(\epsilon_\sigma,\mu_\sigma)$ and $\mathbb{C}^{e}_\sigma={\rm diag}(\epsilon_0,\mu_0)$.
(C2) An asymmetric, material independent choice of coefficients is $\mathbb{C}^{i}_\sigma={\rm diag}(1,0)$ and $\mathbb{C}^{e}_\sigma={\rm diag}(0,1)$ which implies that the electric (magnetic) surface currents are determined by the exterior (interior) solution only.
For good conductors, we have observed fast convergence of the MSE with this choice.
(C3) Finally, we note that the singular choice with $\mathbb{C}^{i}_\sigma+\mathbb{C}^{e}_\sigma=\mathbb{0}$, which we excluded, does not yield a Fredholm integral equation and hence does not permit a MSE. The latter choice has been employed in a computationally intensive boundary element method \cite{reid2013}, implemented in the open-source software SCUFF-EM \cite{scuff}. For the results in this work we shall employ the choice (C1).

\begin{figure}
\centering
\includegraphics[width=.9\linewidth]{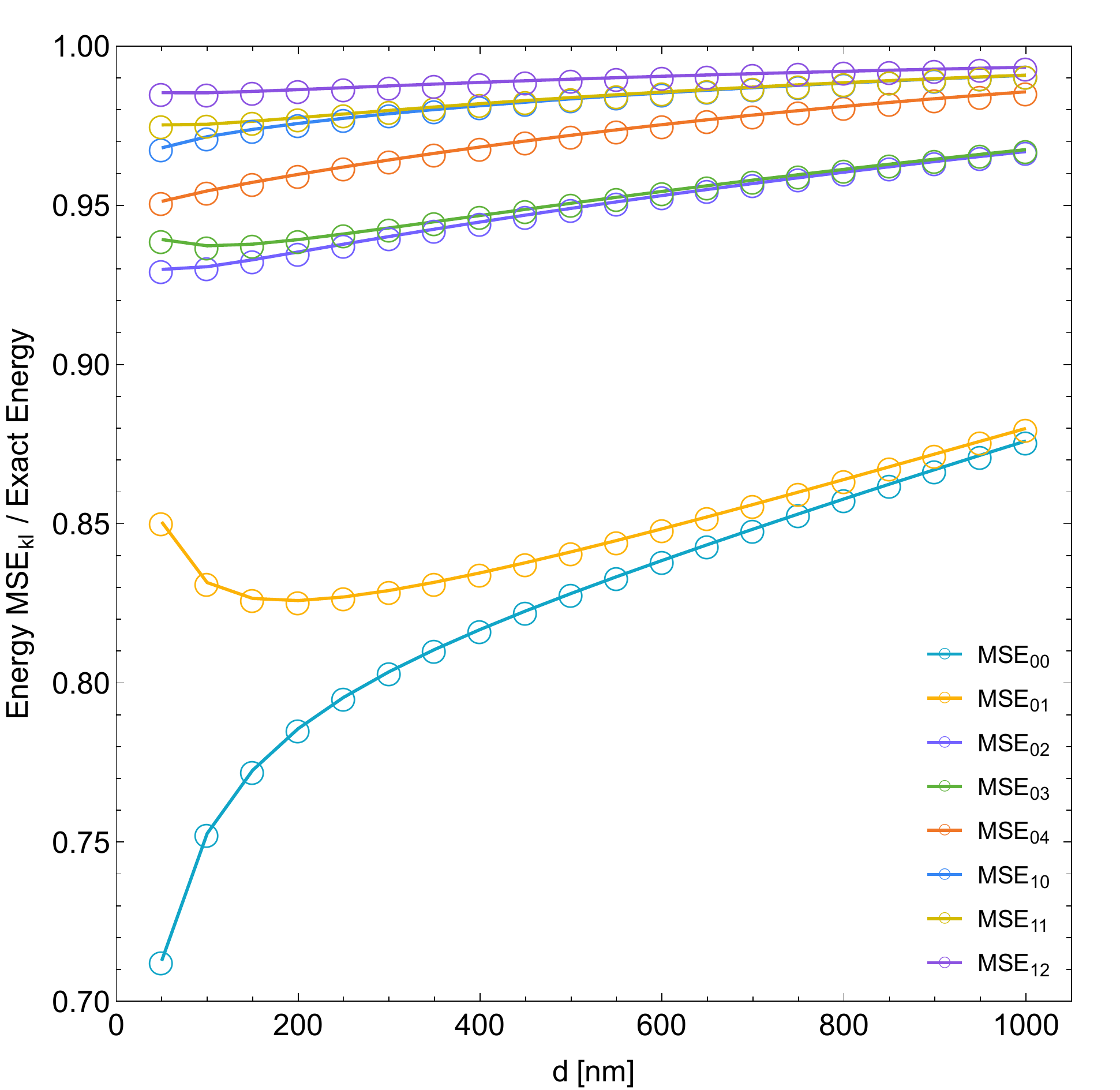}
\caption{\label{fig:2} Different orders of the MSE for the Casimir energy between a plate made of doped silicon and a plate made of gold (see Supplemental Material for details), normalized to the known exact energy. Indices of MSE$_{kl}$ label the number of scatterings between the plates ($2(k+1)$) and within the silicon plate ($l$) (see text for details).}
\end{figure}

{\it Convergence of the expansion in the number of scatterings} --- 
The interaction between two planar and parallel surfaces is determined by their Fresnel coefficients  \cite{lifshitz}. 
The majority of experiments measure forces between gold (Au) and/or doped silicon (Si) surfaces \cite{Wang2021,mohideen,lamoreaux,bressi}, and hence we consider these materials in the following applications of the MSE \footnote{See Supplemental Material for material parameters and technical details, which includes 
additional Refs.~\cite{Maradudin:2007aa,Marachevsky:2021aa,Shanks:1955rq,Buhmann:2006ab,Contreras-Reyes:2010aa}. }. When the SSO $\mathbb{K}_{11}$ describes the scatterings on the Si surface, expansion of the energy in Eq.~(\ref{eq:4}) in this SSO yields MSE approximation to the Casimir interaction. MSE orders are labelled by MSE$_{kl}$ where $2(k+1)$ is the number of scatterings between the surfaces (total number of $\mathbb{K}_{12}$ and $\mathbb{K}_{21}$ operators) and $l$ is the number of single-body scatterings on the Si surface  (number of $\mathbb{K}_{11}$ operators). Fig.~\ref{fig:2} shows the energy for eight different orders of MSE relative to the known exact energy at $T=300$K for surface separations from 100nm to 1$\mu$m. While the lowest order MSE$_{00}$ with {\it no} single-body scattering on the Si surface yields between $70\%$ and $87\%$ of the exact interaction, only 4 scatterings between the surfaces ($k=1$) and 2 single-body scatterings on the Si surface ($l=2$) are required for an accuracy of about $1\%$. This validation example  demonstrates fast convergence of our MSE.

\begin{figure}
\centering
\includegraphics[width=1\linewidth]{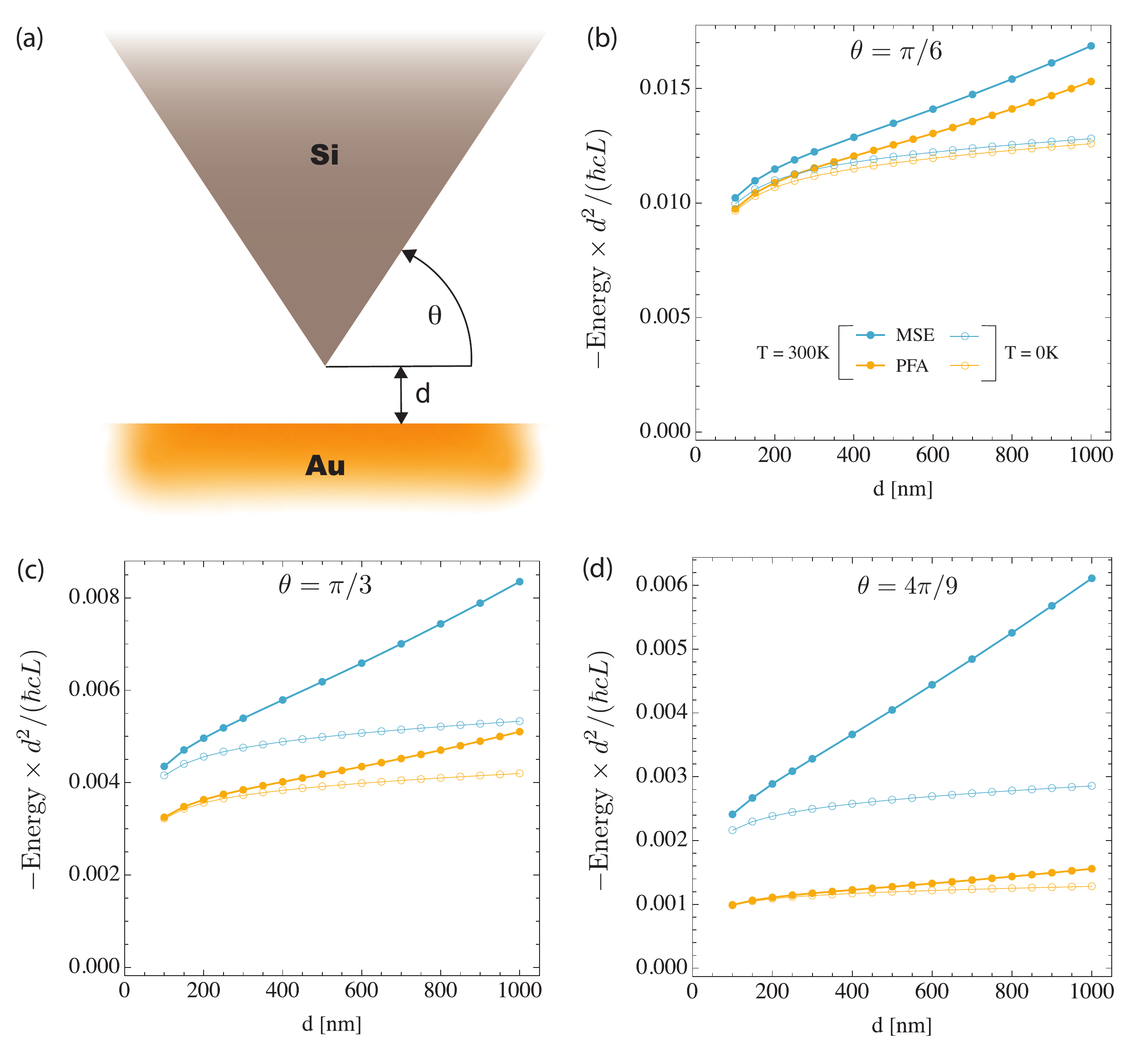}
\caption{\label{fig:3} 
{\bf (a)} Schematic of the wedge-plate configuration with edge-plate separation $d$ and angle $\theta$ between plate and the sides of the wedge. Other panels {\bf (b)} to {\bf (d)} show the Casimir energy of the wedge-plate configuration both at zero temperature and at room temperature ($T=300$K) as a function of the separation $d$, obtained from the multiple scattering expansion (MSE), and the proximity force approximation (PFA), for three different angles $\theta=\pi/6$, $\theta=\pi/3$, and $\theta=4\pi/9$.}
\end{figure}

{\it Surface with sharp edge} --- We apply the MSE to compute the
Casimir interaction between a gold (Au) plate and wedge made of doped silicon (Si), both at zero and room temperature ($T=300$K). The wedge is aligned symmetrically with the normal to the plate and its wings form an angle $\theta$ with the plate, see Fig.~\ref{fig:3}(a). 
To date, the only in principle applicable method to compute Casimir forces for this setup is a fully numerical boundary element discretization (SCUFF-EM) \cite{reid2013}. The T-matrix of a wedge is only known for a perfect metal for which the interaction energy per length $L$ of the edge of the wedge at zero temperature is ${\cal E}/L = - f(\theta) \hbar c / d^2$ at separation $d$, with a dimensionless function $f(\theta)$ which has been computed previously \cite{maghrebi}. Here, our system contains additional length scales such as the thermal wavelength $\lambda_T = \hbar c/k_B T \approx 7.6\,\mu$m at $T=300$K and various length scales characterizing the optical properties of the materials Au and Si. 
The MSE amounts to evaluate Eq.~(\ref{eq:4}) by expanding the logarithm and $[\mathbb{1}-\mathbb{K}_{11}(\theta)]^{-1}$ into a series of iterated integrals of the operator $\Delta\mathbb{K}_{11}(\theta)=\mathbb{K}_{11}(\theta)-\mathbb{K}_{11}(\theta=0)$. Specifically, we perform numerical integrations up to orders MSE$_{04}$ and MSE$_{12}$ for two and four scatterings between plate and wedge, respectively. Fig.~\ref{fig:3} shows the resulting attractive Casimir energy for the three angles $\theta=\pi/6$, $\pi/3$ and $4\pi/9$.
The interaction shows several notable features. First, a comparison with the PFA reveals a breakdown of these approximations even at surface proximity, unless the wings of the wedge open to a plane ($\theta=0$). Indeed, the approximation error increases dramatically from small to large angles, reaching nearly a factor of $5$ for $\theta=4\pi/9$ at the largest studied separation of $1\,\mu$m and $T=300$K. 
Second, we observe anomalously large thermal effects, compared to parallel plates. Fig.~\ref{fig:3} shows that the difference between the interaction at zero and room temperature increases dramatically with $\theta$. 
Thermal effects also modify substantially the 
crossover between the quantum regime at short distances and the classical interaction ${\cal E}_{\rm cl} \sim k_B T L / d$ 
which usually dominate for separations larger than $\lambda_T$. For the Au/Si parallel plates considered before, the classical contribution accounts for $45\%$ of the total interaction at $d=1\,\mu$m. Differently, for the wedge-plate geometry importance of thermal effects emerges already at much shorter separations, with ${\cal E}_{\rm cl}$ contributing  to the total energy 
$73\%$, $79\%$ and $88\%$ with increasing values of the three  angles $\theta$ at $d=1\,\mu$m. 

\begin{figure}
\centering
\includegraphics[width=0.9\linewidth]{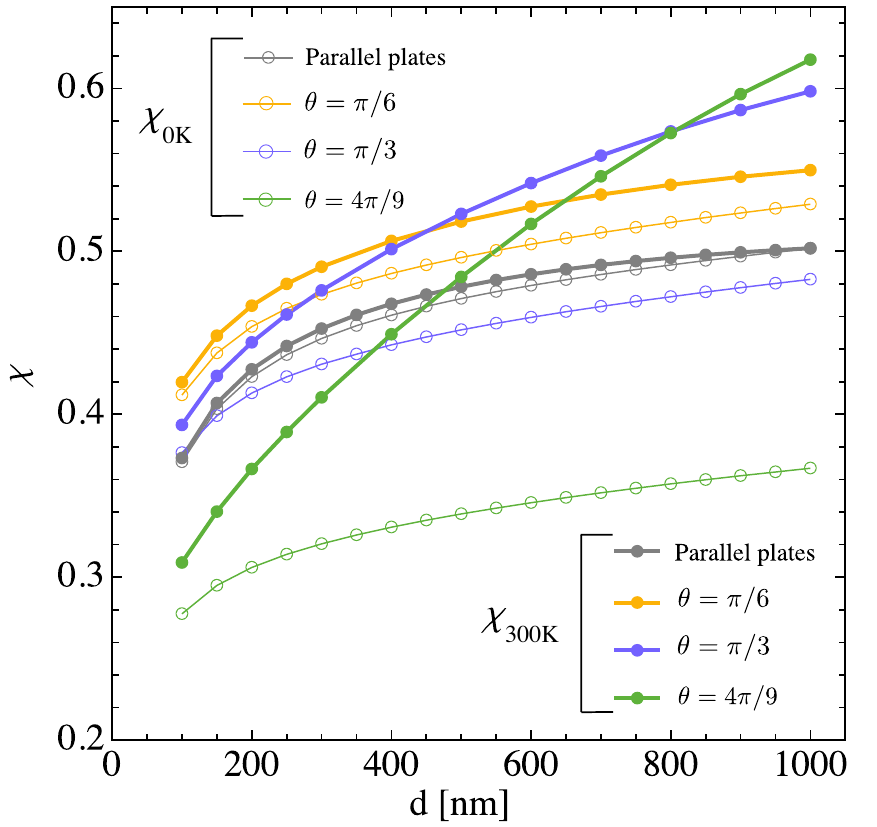}
\caption{\label{fig:4} Influence of material properties and thermal fluctuations on the wedge-plate interaction. Modification factor $\chi$ which compares Casimir energy ${\cal E}_{\rm Au-Si}$ of Au-Si plate-wedge to energy for perfect metal (PM) plate-wedge, both at $T=0$K and $T=300$K. Thin curves with open dots represent $\chi_{0K}={\cal E}_{\rm Au-Si, 0K}/{\cal E}_{\rm PM, 0K}$, and bold curves with filled dots show  $\chi_{300K}={\cal E}_{\rm Au-Si, 300K}/{\cal E}_{\rm PM, 300K}$. See text for details.}
\end{figure}

Finally, it is instructive to compare our findings to  the perfect metal (PM) approximation \cite{maghrebi} which for $T=0$K would be represented in the plots of Fig.~(\ref{fig:4}) by a horizontal line equal to $f(\theta)$. The modification factor $\chi$, defined as the ratio of the energies for  Au-Si and PM, is shown in Fig.~\ref{fig:4}  for both temperatures. A number of interesting observations are made. First, at zero temperature, the effect of finite conductivity is strongest for the sharpest wedge (reduction to $28\%$ of PM energy at $d=100$nm), while for smaller angles $\theta$ the reduction  is surprisingly smaller than for parallel plates.
We interpret this observation by a penetration of high frequency field fluctuations into the material and a resulting effective smoothening of the edge. 
Second, to study the combined effect of thermal and material effects, we also computed the PM energy at $T=300$K, and compared it to the Au-Si energy. We observe that thermal fluctuations reduce the effect of finite conductivity, leading to a larger  factor $\chi$, increasing with separation due to the rapidly increasing importance of thermal effects. 
This comparison to an ideal system shows that the combined effect of conductivity and thermal corrections are strongly shape dependent and important for any experimental study of Casimir forces in complex geometries.

{\it Discussion} --- We demonstrate that within the MSE the interaction can be computed very accurately at all separations, by considering a fairly small number of scatterings. As a result we can provide precise predictions for experiments. For instance, for a silicon wedge with an opening angle of 60 degrees and an edge length of $10\; \mu$m, when placed at a distance of $100$ nm from a gold surface, will experience a normal Casimir force of a few pN, which is within the current experimental resolution \cite{Wang2021}.  

The Casimir-Polder interaction of  an atom and an arbitrarily shaped dielectric surface can be also obtained from our MSE. Particles in cavities  and their  effective polarisabilities  are other important applications \cite{Fiedler:2017aa}.
Several further applications can be envisaged, such as forces in increasingly complex nano-electromechanical systems, actuation forces in systems composed of interlocking bodies, and torques between symmetry-breaking objects. 
Further applications can be found in chemistry and colloidal science, e.g. in the study of collective dispersion interactions in novel complex materials composed of quantum objects  \cite{Fiedler:2023aa}, as well as in the context of phase transitions in the self-aggregation patterns of colloidal suspensions of  lipidic and surfactant nano-particles    controlled by  intermolecular forces \cite{Monduzzi:2019aa}, and also  in biological systems  where the importance of curvature of surfaces and interfaces for functional properties has been demonstrated \cite{Schamberger:aa}.
Importantly, we expect that for media with smaller dielectric contrast, such as those found in soft-matter systems like those listed above, the rate of convergence of the MSE increases even further.
We stress that our approach can be used also to study repulsive forces as the medium surrounding the bodies can be a general dissipative magneto-dielectric medium. Naturally, our description in terms of reflections and free propagations in between along closed paths should reproduce the geometric-optics limit of Casimir interactions \cite{jaffe}, and, more importantly, provide a  framework to systematically compute corrections due to  diffraction. 
 The application of the MSE to non-equilibrium systems (e.g.~heat radiation and transfer)
 requires an analysis of the convergence properties for {\it real} frequencies \cite{balian1977}. Here a suitable choice of coefficient matrices $\mathbb{C}^{i}_\sigma$ is expected to be essential.
In conclusion, our rapidly convergent MSE can provide a powerful tool to delve deeper into Casimir phenomena in submicrometre structures composed of various materials which cannot be understood by simple additive power laws and planar or spherical surface interactions.

\begin{acknowledgments}
Early discussions with B.~Duplantier are acknowledged.
\end{acknowledgments}
\vspace*{-.8cm}

%

\end{document}